\documentclass[reprint,aps,prl,showpacs,floatfix]{revtex4-1}
\usepackage{graphics,epsfig,graphicx}
\usepackage{amsbsy}
\usepackage{amsmath}
\usepackage{bm}
\usepackage{amsfonts}
\usepackage{cancel}
\usepackage{multirow}
\begin{document}

\title{Universality and scaling in the $N$-body sector of Efimov physics} 

\author{M. Gattobigio}
\affiliation{Universit\'e de Nice-Sophia Antipolis, Institut Non-Lin\'eaire de
Nice,  CNRS, 1361 route des Lucioles, 06560 Valbonne, France }
\author{A. Kievsky} 
\affiliation{Istituto Nazionale di Fisica Nucleare, Largo Pontecorvo 3, 56100 Pisa, Italy}

\begin{abstract}
  Universal behaviour has been found inside the window of Efimov physics for
  systems with $N=4,5,6$ particles. Efimov physics refers to the emergence of a
  number of three-body states in systems of identical bosons interacting {\it
  via} a short-range interaction becoming infinite at the verge of binding two
  particles. These Efimov states display a discrete scale invariance symmetry,
  with the scaling factor independent of the microscopic interaction.  Their
  energies in the limit of zero-range interaction can be parametrized, as a
  function of the scattering length, by a universal function. We have found,
  using the  form of finite-range scaling introduced in [A.~Kievsky and 
  M.~Gattobigio, Phys. Rev A {\bf 87}, 052719 (2013)], that the same universal
  function can be used to parametrize the ground- and excited-energy of $N\le6$
  systems inside the Efimov-physics window. Moreover, we show that the same
  finite-scale analysis reconciles experimental measurements of three-body
  binding energies with the universal theory.
 \end{abstract}
\maketitle

Universality is one of the concepts that have attracted physicists
along the years. Different systems, having even different energy scales, share
common behaviours. The most celebrated example of universality comes from the
investigation of critical
phenomena~\cite{wilson:1983_rev.mod.phys.,fisher:1998_rev.mod.phys.}: at the
critical point, materials that are governed by different microscopic
interactions share the same macroscopic laws, for instance the same critical
exponents. The theoretical framework to understand universality has been
provided by the renormalization group (RG); the critical point is mapped onto a
fixed point of a dynamical system, the RG flow, whose phase space is represented
by Hamiltonians.  At the critical point the systems have scale-invariant (SI)
symmetry, forcing all of the observables to be exponential functions of the control
parameter. A consequence of SI symmetry is the scaling of the observables: for
different materials, in the same class of universality, a selected observable can
be represented as a function of the control parameter and, provided that both
the observable and the control parameter are scaled by some material-dependent
factor, all representations collapse onto a single universal
curve~\cite{stanley:1999_rev.mod.phys.}.

More recently, a new kind of universality has captured the interest of
physicists, namely the Efimov
effect~\cite{efimov:1970_phys.lett.b,efimov:1971_sov.j.nucl.phys.}. A system of
three identical bosons interacting {\it via} two-body short-range interaction
whose strength is tuned, by scientists or by nature, to the verge of binding the
two particle subsystem, exhibits the appearance of an infinite tower of
three-particle bound states, whose energies accumulate to zero. Moreover, the
ratio between the energies of two consecutive states is constant and
independent of the very nature of the interaction; this last property points out
to the emergence of a discrete scale invariance (DSI) symmetry (for a complete review,
see Ref.~\cite{braaten:2006_physicsreports}). 

Even this example of universality has found in the RG its theoretical 
framework. Systems sharing Efimov effect are mapped onto a limit cycle of the
RG flow, where they manifest the emergence of DSI. In turn, DSI implies that all of the
observables are log-periodic function of the control
parameter~\cite{sornette:1998_physicsreports}, and this property is what
characterizes the Efimov physics, of which Efimov effect is an example.
The limit cycle implies the emergence of a new dimensional quantity,
which in the case of Efimov physics is known as the three-body parameter.  Strictly
speaking, the DSI is an exact symmetry for systems with zero-range interaction,
or equivalently in the scaling limit;
for real systems, which posses an interaction with finite range $r_0$,
there are deviations from DSI called finite-range effects. 

Atomic physics, and more precisely experiments using ultracold-alkali atoms, has
recently (re)sparked the interest in Efimov physics~\cite{kraemer:2006_nature}.
At present, several different experimental groups have observed the Efimov
effect in alkali
systems~\cite{ferlaino:2011_few-bodysyst.,machtey:2012_phys.rev.lett.,roy:2013_phys.rev.lett.,dyke:2013_arxiv:1302.0281[cond-mat.quant-gas]},
where the key point has been the scientists' ability to change the two-body
scattering length $a$ by means of Fano-Feshbach resonances. In fact, the
theory predicts how observables change as a function of the control parameter,
$\kappa_*a$, which is proportional to the scattering length, making the tuning of $a$
crucial to test theory's predictions. In particular, the Efimov equation for
the three-body binding energies $E_3^n$ can be expressed in a parametric form as
follow~\cite{braaten:2006_physicsreports}
\begin{equation}
  \begin{gathered}
    E_3^n/(\hbar^2/m a^2) = \tan^2\xi \\
    \kappa_*a = \text{e}^{(n-n^*)\pi/s_0} 
    \frac{\text{e}^{-\Delta(\xi)/2s_0}}{\cos\xi}\,,
  \end{gathered}
  \label{eq:energyzr}
\end{equation}
with $\Delta(\xi)$ a universal function whose parametrization can be
found in Ref.~\cite{braaten:2006_physicsreports}, $s_0=1.00624$, and
$\kappa_*$ the emergent three-body parameter which  gives the energy 
$\hbar^2 \kappa_*^2/m$ for $n=n^*$ at the unitary limit $1/a=0$. 

The ability of tuning $a$, has allowed the different experimental groups to
measure the value of the scattering length $a_-$ at which the three-body bound
state disappears into the continuum ($\xi\rightarrow -\pi$). From
Eq.~({\ref{eq:energyzr}) we see that measuring $a_-= -
\text{e}^{-\Delta(\pi)/2s_0}/\kappa_*\approx
-1.56/\kappa_*$~\cite{esry:1999_phys.rev.lett.,bedaque:2000_phys.rev.lett.},   is a way to
measure the three-body parameter $\kappa_*$, which in principle should be
different for different systems. However, it has been experimental
found~\cite{ferlaino:2011_few-bodysyst.,machtey:2012_phys.rev.lett.,roy:2013_phys.rev.lett.,dyke:2013_arxiv:1302.0281[cond-mat.quant-gas]},
and theoretically
justified~\cite{wang:2012_phys.rev.lett.,naidon:2012_arxiv:1208.3912[cond-mat.quant-gas]},
that in the class of alkali atoms $a_-/\ell\approx -9.5$, with $\ell$ the van
der Waals length; a universality inside universality. Recently the same behavior
has been seen in a gas of $\,^4$He atoms~\cite{knoop:2012_phys.rev.a}.

Eq.~({\ref{eq:energyzr}), as well as the parametrization of $\Delta(\xi)$
have been derived in the scaling limit, where the DSI
is exact. Experiments and calculations made for real systems deal with 
finite-range interactions, $r_0\neq 0$, and for this reason finite-range 
corrections have to be
considered~\cite{efimov:1991_phys.rev.c,thogersen:2008_phys.rev.a,dincao:2009_j.phys.b,platter:2009_phys.rev.a}.
In a recent paper~\cite{kievsky:2013_phys.rev.a}, the authors have observed
in which manner finite-range corrections manifest in numerical calculations using
potential models:

(i) There are corrections coming from the two-body sector which can be
taken into account by substituting $a_B$ for $a$, defined by
$E_2=\hbar^2/m a_B^2$, with $E_2$ the two-body binding energy if $a>0$, or the two-body
virtual-state energy in the opposite case, $a<0$~\cite{ma:1953_rev.mod.phys.}.
One simple way to obtain the
virtual-state energy is looking for the poles of the two-body S-matrix using a Pad\`e
approximation, as shown in Ref.~\cite{rakityansky:2007_j.phys.a:math.theor.}.
It should be noticed that in the zero-range limit $a_B\rightarrow a$.

(ii) The finite-range
corrections enter as a shift in the control parameter $\kappa_*a_B$. 
The value of the shift depends on the observable under investigation. 

For instance, in the case of three-body binding energies, (i) and 
(ii) applied to Eq.~(\ref{eq:energyzr}) give

\begin{equation}
  \begin{gathered}
    E_3^n/E_2 = \tan^2\xi \\
    \kappa^3_na_B + \Gamma_n^3 =
    \frac{\text{e}^{-\Delta(\xi)/2s_0}}{\cos\xi} \,,
  \end{gathered}
  \label{eq:energyRange}
\end{equation}
where we have defined $\kappa_n^3 = \kappa_*\text{e}^{(n-n^*)\pi/s_0}$, 
and introduced the shifts $\Gamma_n^3$. 
In Ref.~\cite{kievsky:2013_phys.rev.a} the authors have
shown that this type of correction appears in the energy spectrum, in atom-dimer scattering length 
and in the effective range function of three boson atoms. Moreover, in
Ref.~\cite{garrido:2013_phys.rev.a} it has been shown that the shift appears in 
recombination rate of three atoms close to threshold too. 

  \begin{figure}
    \includegraphics[width=\linewidth]{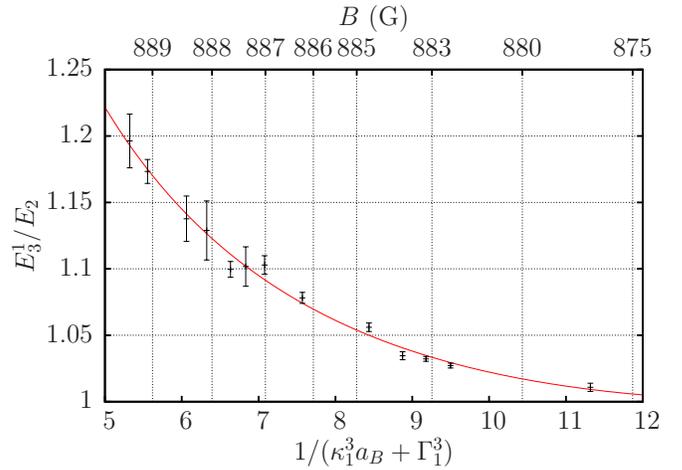}
    \caption{(color online). The experimental data on
      $\ ^7$Li~\cite{machtey:2012_phys.rev.lett.}, in the form of
      ratio between the three-body binding energy $E_3^1$ and the two-body
      binding energy $E_2$, as a function of $1/(\kappa_1^3a_B+ \Gamma_1^3)$ with the
      values of the three-body parameters given in the text. The
      solid curve represents the prediction of the universal function,
      Eq.~(\ref{eq:energyRange}). In the upper abscissa the
      magnetic field from Ref.~\cite{gross:2011_comptesrendusphysique} is given.}
  \label{fig:data}
  \end{figure}

Our finite-range analysis can be applied to describe experimental data.
In Fig.~\ref{fig:data} we report the experimental three-body binding energies
measured in $\ ^7$Li~\cite{machtey:2012_phys.rev.lett.} and
for reference the corresponding magnetic field~\cite{gross:2011_comptesrendusphysique}.
Using Eq.~(\ref{eq:energyRange}) with the values of the two three-body parameters
$\Gamma_1^3=4.95\times 10^{-2}$ and $\kappa^3_1 = 1.61\times 10^{-4}\,a_0$ the
experimental point collapse on the universal curve (solid line).
A more extended analysis is underway.

To explain the origin of this form of finite-range correction, we refer to the
original derivation~\cite{efimov:1971_sov.j.nucl.phys.} of
Eq.~(\ref{eq:energyzr}) and to the parametrization of the universal phase
$\Delta(\xi)$, for instance in Ref.~{\cite{braaten:2006_physicsreports}.
In the zero-range limit $r_0=0$ the adiabatic approximation is exact, and 
the three-body problem is equivalent to a single Schr{\"o}dinger equation
in a scale-invariant $1/R^2$ potential, where
$R^2\propto r_{12}^2+r_{13}^2+r^2_{23}$ is the hyperradius; 
Eq.~(\ref{eq:energyzr}) has been derived by matching the scale-invariant phase
shift $\Delta(\xi)$ originating from the long-range physics to the
scaling-violating phase shift originating from the short-range physics (see
Eq.(193) of Ref.~\cite{braaten:2006_physicsreports}). The short-range physics
can be encoded in a scale-violating momentum $\Lambda_0$, see Eq.~(147) of
\cite{braaten:2006_physicsreports}, and the parametrization, for a zero-range
theory, of $\Delta(\xi)$ is such that $\Lambda_0 = \kappa_*$.
Now, when we consider a finite-range system, $r_0\neq 0$, the lowest adiabatic potential 
is coupled to the other adiabatic potentials: for instance, it has been demonstrated by
Efimov~\cite{efimov:1991_phys.rev.c} that the coupling can be taken into account by a correction $\sim r_0/R^3$ on the
lowest potential. This means that, keeping the same parametrization of
$\Delta(\xi)$, the relation between $\Lambda_0$ and  $\kappa_*$ is modified, and 
at the first order we can expect
\begin{equation}
  \Lambda_0 \simeq \kappa_*\left(1+{\cal A}\, \frac{r_0}{a_B}\right)\,,
  \label{eq:finiteCorrection}
\end{equation}
which gives the shift $\Gamma^3_{n^*} = {\cal A}\,\kappa_*r_0$. The constant
${\cal A}$ is expected to take natural values.

In this work we extend the application of the modifications to the zero-range
theory in order to analyze the ground- and
excited-binding energy of $N$-body systems obtained by numerical calculations
inside the window of Efimov physics.  The Efimov effect is strictly related to
the $N=3$ system, but one can try to investigate if and how Efimov physics
affects $N>3$ sectors. Some seminal-theoretical
studies~\cite{platter:2004_phys.rev.a,von_stecher:2009_natphys,deltuva:2013_few-bodysyst},
and subsequent experimental investigation~\cite{ferlaino:2009_phys.rev.lett.},   have
demonstrated that for each trimer belonging to the Efimov tower there are two
attached four-body
states.
This property has also been observed in
$N=5,6$~\cite{gattobigio:2011_phys.rev.a,von_stecher:2011_phys.rev.lett.,gattobigio:2012_phys.rev.a}:
there are two attached five-body states to the four-body ground state and there
are two attached six-body states to the five-body ground state. These states
have been characterized by measuring ratios between energies close to the
unitary limit, and these ratios have been found to be universal. Moreover, their
stability has been analyzed along the Efimov plane in wide region of the angle
$\xi$ \cite{gattobigio:2013_few-bodysyst}.

We want to make a step forward showing that 
the  three-body equation, Eq.~({\ref{eq:energyRange}), can be modified
to predict $N$-body ground- and excited-state energies $E^0_N$ and $E_N^1$.
Even though our calculations have been done up to $N=6$, a clear indication of
validity for generic $N$ can be inferred. We have solved the Schr\"odinger
equation using two different potential: (i) the first is
an attractive two-body gaussian (TBG) potential
$V(r)=V_0 \,\, {\rm e}^{-r^2/r_0^2}$, 
where  $r_0$ is the range of the potential and $V_0<0$ the strength that can be
modified in order to tune the scattering length inside the Efimov window.
This kind of potential has been previous used to investigate clusters of 
$\
^4$He~\cite{kievsky:2011_few-bodysyst.,gattobigio:2012_phys.rev.a,kievsky:2013_phys.rev.a,garrido:2013_phys.rev.a},
and some numerical results, used in this work, has been previous given
in Ref.~\cite{gattobigio:2012_phys.rev.a}. 
(ii)~The second is a P\"oschl-Teller (PT)
potential~\cite{kruppa:2001_phys.rev.c}
\begin{equation}
  V(r) = -
  \frac{\hbar^2}{mb^2}\,\frac{2(1+C)}{\cosh^2(r/b)}\,,
  \label{eq:poschlTeller}
\end{equation}
where the dimensionless parameter $C$ can be varied to change the scattering
length.
The solution of the $N$-body Schr{\"o}dinger equation has been found using  the
non-symmetrized hyperspherical harmonic (NSHH) expansion method with the
technique recently developed by the authors in
Refs.~\cite{gattobigio:2009_phys.rev.a,gattobigio:2009_few-bodysyst.,
gattobigio:2011_phys.rev.c,gattobigio:2011_phys.rev.a}

\begin{figure}[h]
    \begin{center}
      \includegraphics[width=\linewidth]{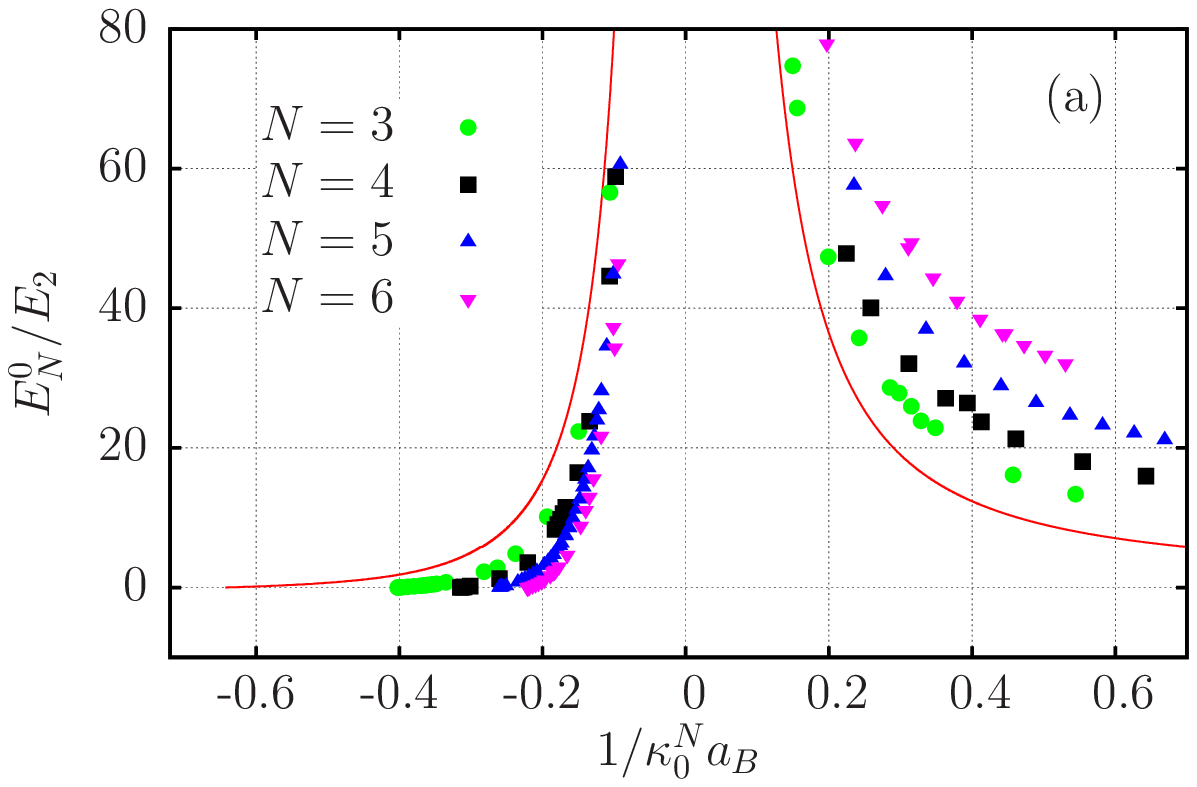}
      \includegraphics[width=\linewidth]{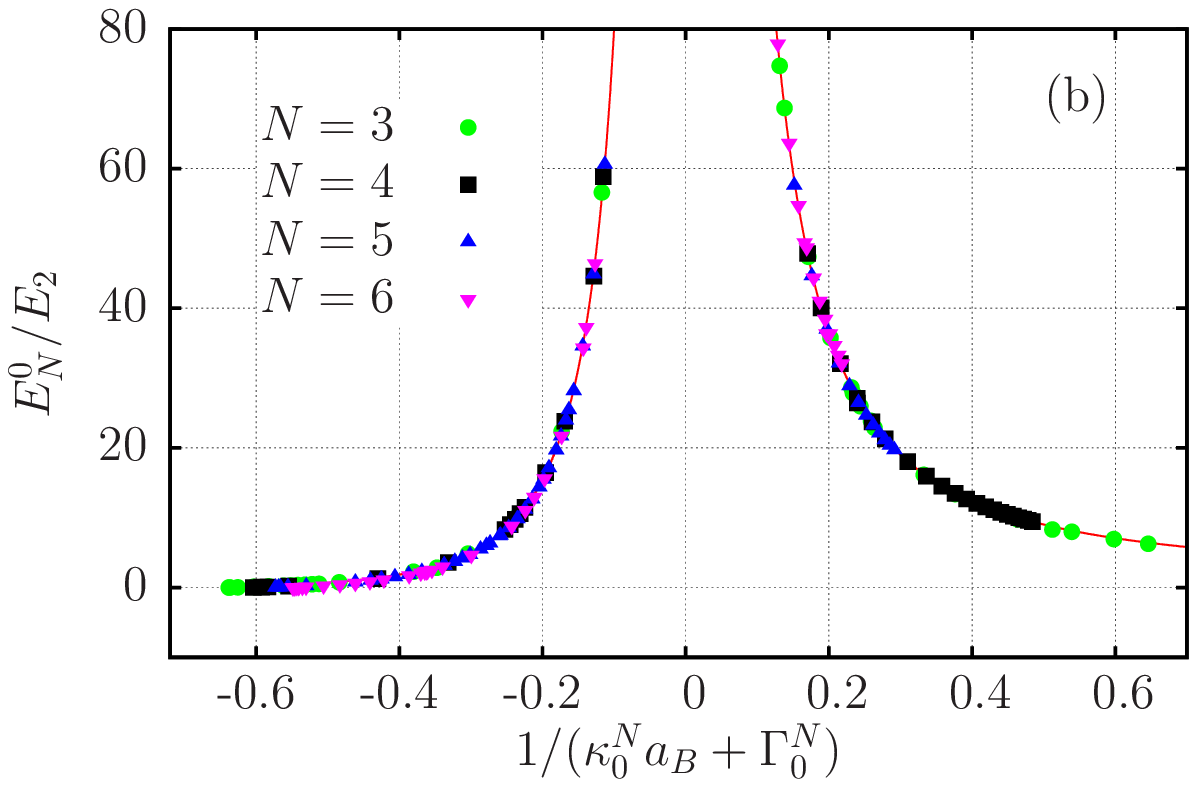}
    \end{center}
    \caption{(color online). The ground-state N-body binding-energy
    $E^0_N$ in units of $E_2$ as a function of (a)  $1/\kappa_0^N a_B$, 
  and (b) $1/(\kappa_0^N a_B +\Gamma_0^N$).  $E_2$ is the two-body binding
  energy for $a>0$, and the two-body virtual state energy for $a<0$.
  In panel (b) the ground-state energies in terms of the scaled
  variable collapse onto the zero-range curve (solid line).}
  \label{fig:universalityZoom}
  \end{figure}

In Fig.~\ref{fig:universalityZoom}(a) we show selected results for the
$N=3,4,5,6$ ground-state binding energies. The $N$-body ground-state  binding
energies $E_N^0$ are divided by $E_2$, which is the two-body binding energy for
$a>0$ or the virtual-state energy for $a<0$. These ratios are given as a
function of the inverse of the control parameter $\kappa_0^N a_B$. The parameter
$\kappa_0^N$ is fixed by the $N$-body ground-state binding energy $E_N^0 =
\hbar^2(\kappa_0^N)^2/m$ calculated at the unitary limit $1/a=0$. The
corresponding values and some relevant ratios are given in
Table~\ref{tab:kappa}. The solid curve represents the result of the $N=3$
zero-range theory given in  Eq.~(\ref{eq:energyzr}) for $n=n^*$. 
The results for the different clusters have been obtained
using both TBG and PT potentials.
In the figure only a subset of the numerical data is
shown in order to better appreciate the trend.

In Fig.~\ref{fig:universalityZoom}(b) the same data are shown, but this time the
control parameter, $\kappa_0^N a_B$,
has been shifted by a quantity $\Gamma^N_0$, different for each
particle sector. As a remarkable result, the different sets of data
collapse on the three-body zero-range universal curve.
This is very reminiscent of the scaling property in critical
phenomena~\cite{stanley:1999_rev.mod.phys.}. 
In our case we have a $N$-dependent parameter, $\kappa_0^N$, that fixes the
scale of the system and, in this respect, we refer here to it as a scaling 
parameter. Furthermore an $N$-dependent parameter, $\Gamma_0^N$ 
appears to take into account finite-range corrections. In this respect we refer
to it as a finite-range scaling parameter.

It should be noticed that the values of $\kappa_0^N$ has been obtained from our
data, and in doing so we have included some range corrections into these
quantities. As it is well known, the lower energy states, as those considered
here, have some dependence on the form of the potential. This dependence
decreases in higher level states~\cite{deltuva:2013_few-bodysyst}. However, we
want to emphasize that $\kappa_0^N$ are not new $N$-body parameters in the same
sense as the emergent three-body parameter $\kappa_*$. In the present treatment,
where we only use two-body interaction (eventually, we could have also
added  three-body
interactions~\cite{gattobigio:2012_phys.rev.a}), there are not such a thing as four-, five-, and six-body
parameters; in the scaling limit all their values are fixed by the three-body
parameter $\kappa_*$, as discussed below (for a different point of view
see Ref.~\cite{hadizadeh:2011_phys.rev.lett.}).

\begin{figure}[h]
    \includegraphics[width=\linewidth]{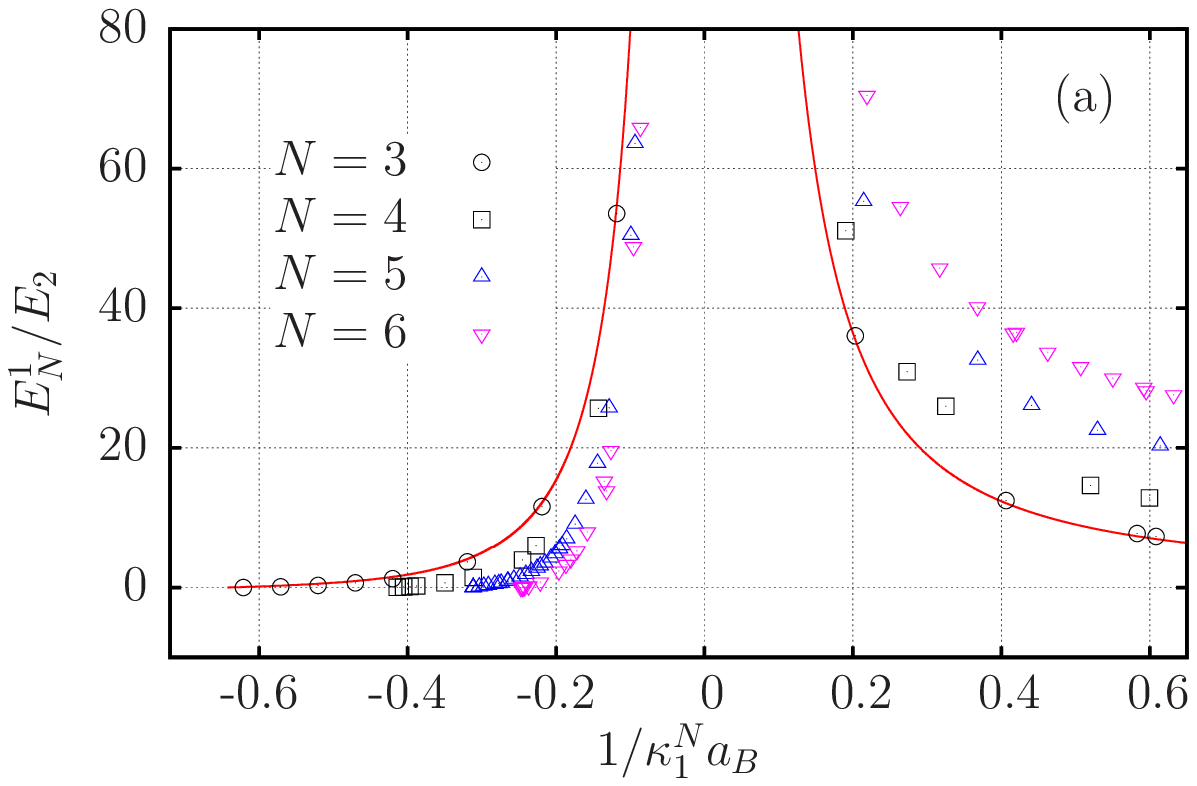}
    \includegraphics[width=\linewidth]{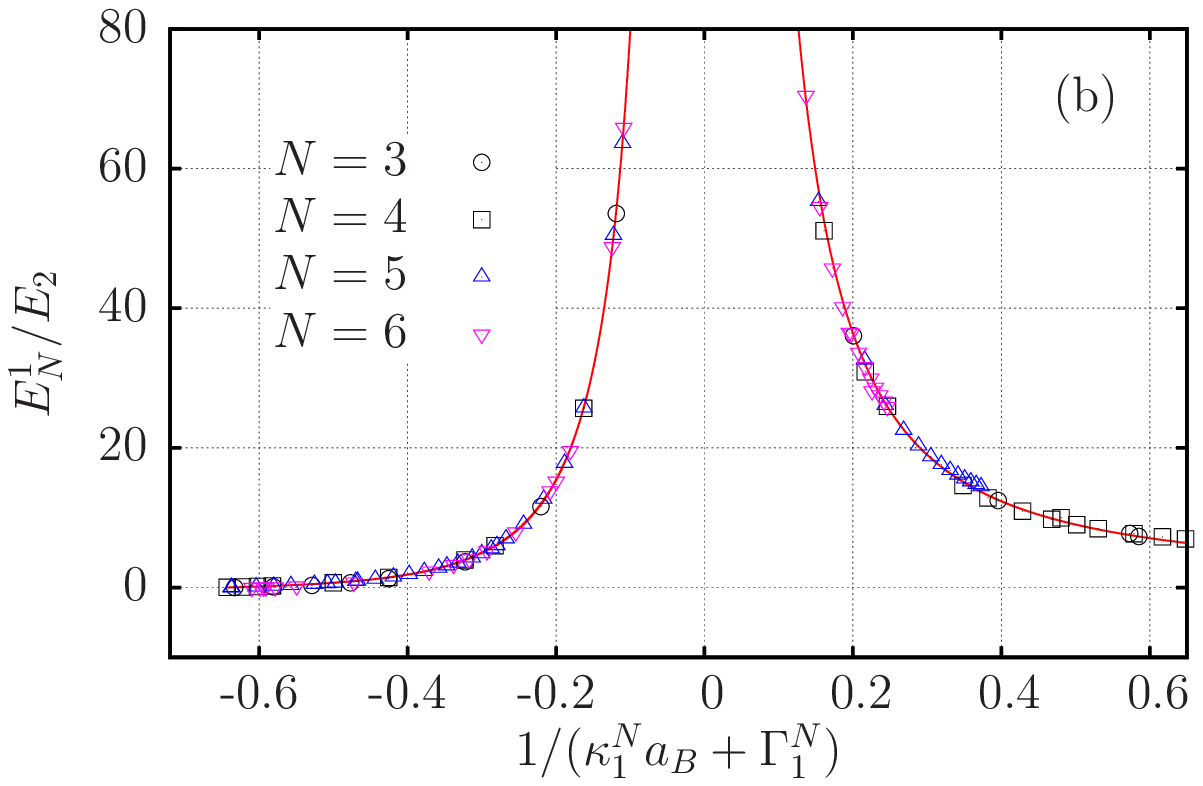}
    \caption{(color online). The excited-state N-body binding-energy
    $E^1_N$ in units of $E_2$ as a function of (a)  $1/\kappa_1^N a_B$, 
  and (b) $1/(\kappa_1^N a_B +\Gamma_1^N$).  $E_2$ is the two-body binding
  energy for $a>0$, and the two-body virtual state for $a<0$.
  In panel (b) the excited-state energies in terms of the scaled variable collapse
onto the zero-range curve (solid line).}
  \label{fig:universalityExcitedZoom}
  \end{figure}

In Fig.~\ref{fig:universalityExcitedZoom}(a) we show our calculations for the
$N$-body excited states $E^1_N$. 
We report the ratios $E_N^1/E_2$, where 
$E_2$ is still either the two-body binding energy for $a>0$ or the virtual-state
energy for $a<0$, as a function of the inverse of the control parameter
$\kappa_1^N a_B$. As for the ground states, the parameters $\kappa_1^N$ are 
fixed by the excited-binding energy 
$E^1_N = \hbar^2(\kappa_1^N)^2/m$ at the unitary limit.
The solid line
shows the universal function.
Again, we want to stress that
they are not new $N$-body parameters, but they are fixed by the value of
$\kappa_*$. As before,
$\kappa_1^N$ has 
some range 
corrections which
can be estimated in the case of $N=3$
from Table~\ref{tab:kappa}. The zero-range theory imposes
$\kappa_0^3/\kappa_1^3\approx 22.7$ whereas 
we found $\approx 23.0$ and $\approx 22.4$ using TBG and PT potentials
respectively. 

In Fig.~\ref{fig:universalityExcitedZoom}(b) our data sets are shown with
a shift in the control variable $\kappa_1^N a_B$ by
a $N$-dependent quantity  $\Gamma_1^N$. As for the ground states, the 
excited states collapse on the universal curve too, pointing out to the 
emergence of a common universal behaviour in the $N$-boson system.

  \begin{figure}[h]
    \includegraphics[width=\linewidth]{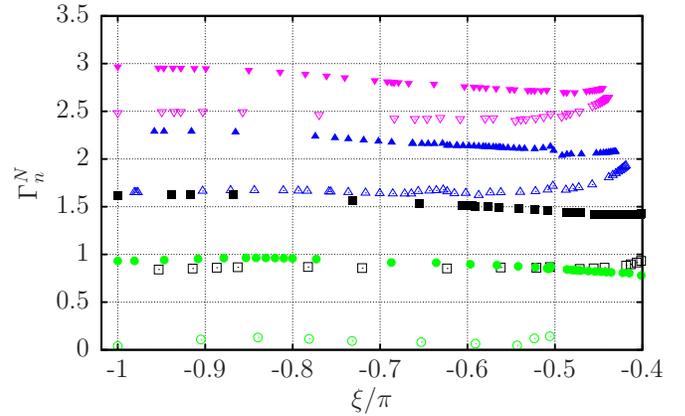}
    \caption{The finite-range scaling parameters $\Gamma_n^N$ as a function of the 
     angle $\xi$ derived from Eq.~(\ref{eq:energy}). The symbols are the same as 
     in Figs.~\ref{fig:universalityZoom} and \ref{fig:universalityExcitedZoom}.}
  \label{fig:deltas}
  \end{figure}

Our numerical findings can be summarized in a modified version of 
Eq.~(\ref{eq:energyzr}). We propose
\begin{equation}
  \begin{gathered}
    E_N^n/E_2 = \tan^2\xi \\
    \kappa_n^N a_B + \Gamma_n^N =  
    \frac{\text{e}^{-\Delta(\xi)/2s_0}}{\cos\xi}\,.
  \end{gathered}
  \label{eq:energy}
\end{equation}
where the function $\Delta(\xi)$ is universal and it is determined by the 
three-body physics. The above equation, valid for general $N$, shows the same 
universal character of the three-boson system and, due to the DSI, with
the same universal function $\Delta(\xi)$. The parameter $\kappa_n^N$ appears
as a scale parameter and the shift $\Gamma_n^N$ is a finite-range scale
parameter introduced to take into account finite-range corrections.
The introduction of the shifts $\Gamma_n^N$ is probably a first-order correction
of finite-range effects. In fact, we can use Eq.~(\ref{eq:energy}) to see that 
a small dependence on the parameter $\xi$ still remains. This is illustrated in 
Fig.~\ref{fig:deltas} in which $\Gamma_n^N$
is obtained by subtracting to the universal term 
$\text{e}^{-\Delta(\xi)/2s_0}/{\cos\xi}$ the computed value $\kappa_n^N a_B$
at the corresponding values of the angle $\xi$.

Finally, we want to comment on 
the scaling parameters $\kappa_n^N$. In the zero-range limit, their values 
are fixed by the three-body parameter $\kappa_*$. For instance, in $N=3$
we have defined  $\kappa^3_n = \text{e}^{-(n-n^*)\pi/s_0}\kappa_*$, and 
in the four-body sector an accurate study gives $\kappa^4_0 =
2.147\kappa_0^3$~\cite{deltuva:2013_few-bodysyst}. In Table~\ref{tab:kappa}
we report our values for TBG potential~\footnote{In the Supplemental Material we 
report the details of the calculations using PT potential together with an 
analysis of data of Ref.
\cite{von_stecher:2010_j.phys.b:at.mol.opt.phys.}.}, and when available, 
the zero-range-limit values. From the table we can deduce
a linear relation between the ground states that can be approximated as
\begin{equation}
 \frac{\kappa_0^N}{\kappa_0^3} = 1+ (N-3) (\frac{\kappa_0^4}{\kappa_0^3} -1)\,.
\end{equation}
In the scaling limit, using the universal value of $\kappa_0^4/\kappa_0^3$, this
relation reduces to $\kappa_0^N/\kappa_0^3 = 1+ 1.147 (N-3)$. 
The linear relation with $N$ can also been seen in
Refs.~\cite{von_stecher:2010_j.phys.b:at.mol.opt.phys.,hanna:2006_phys.rev.a}.

\begin{table}[h]
  \begin{tabular}{ c | c  c c c }
    \hline
    \hline
     & $N=3$  & $N=4$ & $N=5$ & $N=6$  \\
     \hline
     $\kappa_0^Nr_0$  & $4.88\cdot 10^{-1}$  &  $1.18$  &
         $1.96 $  & $2.77$  \\    
     $\kappa_1^Nr_0$  & $2.12\cdot 10^{-2}$ &$5.11\cdot 10^{-1}$ & $1.24$ & $2.07$\\    
     $\kappa_0^N/\kappa_1^{N}$    & 23.0 (22.7\cite{efimov:1970_phys.lett.b})  &  2.31  & 1.58 & 1.34\\    
     $\kappa_0^N/\kappa_0^{N-1}$  & -      &  2.42
             (2.147\cite{deltuva:2013_few-bodysyst})  & 1.66 & 1.41\\    
     $\kappa_1^N/\kappa_0^{N-1}$  & -      &  1.05
     (1.001\cite{deltuva:2013_few-bodysyst})  & 1.05 & 1.06\\
     $\kappa_1^N/\kappa_1^{N-1}$  & -      &  24.1  & 2.43 & 1.67 \\    
    \hline
  \end{tabular}
  \caption{We report the parameters $\kappa_n^N$, in unit of $r_0$, and selected
  ratios for the TBG potential.  When available, we report 
the ratios at the scaling limit between parenthesis.}
  \label{tab:kappa}
\end{table}

To summarize, we have extended to the $N$-boson systems the concept of
universality inside the Efimov window. By introducing $N$-body scaling
parameters $\kappa_n^N$ and finite-range corrections, $\Gamma_n^N$ and $a_B$, we
have demonstrated that scaled ground- and excited-state energies of systems up
to (at least) $N=6$ collapse over the same universal curve, described by the
universal function appearing in Eq.(\ref{eq:energy}) (for the ensemble of all
the calculated data see the Supplemental material~\cite{Note1}).

As an application, we have shown that our finite-range analysis reconciles
experimental measurements of trimer-binding energies on $\
^7$Li~\cite{machtey:2012_phys.rev.lett.} with the universal theory, showing the
collapse of the data on the universal curve.

\begin{acknowledgments}
We grateful acknowledge Prof. Lev Khaykovich for providing us with the
experimental data of
Refs.~\cite{machtey:2012_phys.rev.lett.,gross:2011_comptesrendusphysique}.
\end{acknowledgments}


%

%
  \renewcommand{\theequation}{S\arabic{equation}}
  \setcounter{equation}{0}
  \renewcommand{\thefigure}{S\arabic{figure}}
  \setcounter{figure}{0}
  \renewcommand{\thetable}{S\arabic{table}}
  \setcounter{table}{0} 
  \section{Supplemental material} 

\subsection{Calculation with P\"oschl-Teller potential}
The  P\"oschl-Teller (PT) potential has the following form
\begin{equation}
  V_\epsilon(r) = -
  \frac{\hbar^2}{mb^2}\,\frac{2(1+C\epsilon)}{\epsilon^2\cosh^2(r/b\epsilon)}\,,
  \label{eq:poschlTellerSp}
\end{equation}
where $b$ is a length scale, $m$ the mass of the identical particles, and $C$
and $\epsilon$ are numbers. The potential is a local-potential representation of
the contact interaction in the limit $\epsilon\rightarrow
0$~\cite{kruppa:2001_phys.rev.c}.

\begin{figure}[h]
    \begin{center}
      \includegraphics[width=\linewidth]{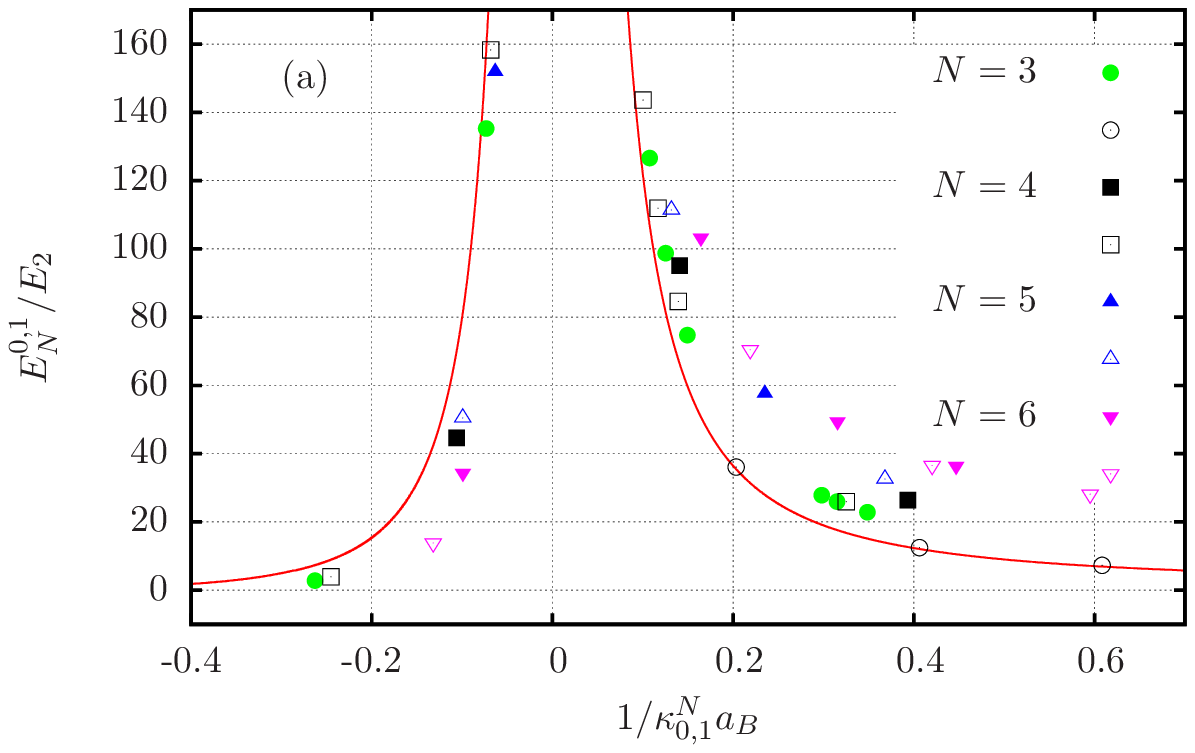}
      \includegraphics[width=\linewidth]{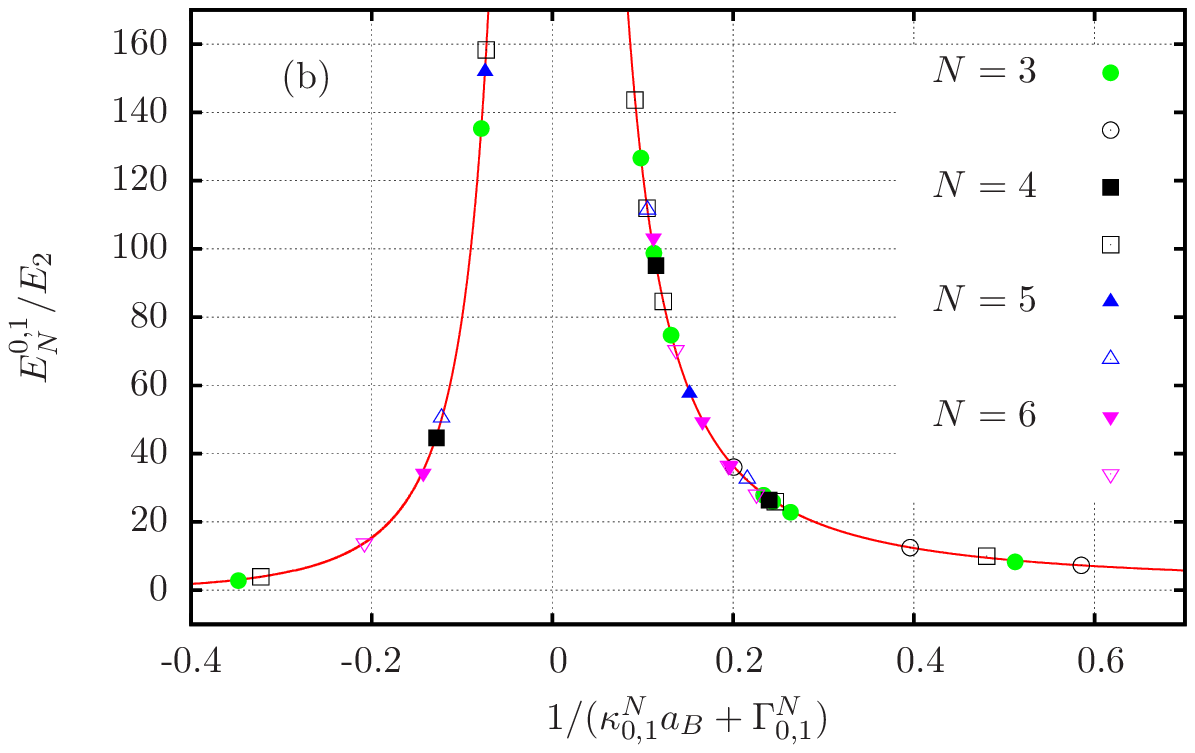}
    \end{center}
    \caption{The ground- and excited-state N-body binding energies $E_N^{0,1}$ of 
      P\"osch-Teller potential, Eq.~(\ref{eq:poschlTellerSp}), 
       in units of $E_2$ as a function of (a)  $1/\kappa_{0,1}^N a_B$, 
      and (b) $1/(\kappa_{0,1}^N a_B +\Gamma_{0,1}^N$).  $E_2$ is the two-body binding
  energy for $a>0$, and the two-body virtual state energy for $a<0$.
  In panel (b) the ground- and excited-state energies in terms of the scaled
  variable collapse onto the zero-range curve (solid line).}
  \label{fig:universality_PT}
  \end{figure}

We made our calculations setting $\epsilon=1$ and varying $C$ in order to change
the scattering length.  We calculated both ground- and excited-state energies
for $N=3,4,5,6$, and 
a selection of our results are shown in
Fig.~\ref{fig:universality_PT}; on Fig.~\ref{fig:universality_PT}(a) we present
our calculation without shift, while data in Fig.~\ref{fig:universality_PT}(b) 
are shifted and we see that they collapse over the universal curve.
Moreover, in Table~\ref{tab:datiTP} 
we report 
a summary of the shifts and the energies at the unitary limit for PT potential. 
\renewcommand{\arraystretch}{1.5}
\begin{table}[h]
\begin{center}
  \begin{tabular}{ c | c  c c c }
    \hline
    \hline
     & $N=3$  & $N=4$ & $N=5$ & $N=6$  \\
     \hline
     $\kappa_0^N b$  &  0.3668 & 0.9088 & 1.521 & 2.175\\
     $\kappa_1^N b$  &         & 0.3934 & 0.971 & 1.633\\
     $\kappa_0^N/\kappa_1^{N}$ &   & 2.31 & 1.58 & 1.33\\
     $\kappa_0^N/\kappa_0^{N-1}$  & -  & 2.48  & 1.68 & 1.43\\
     $\kappa_1^N/\kappa_0^{N-1}$  & -  & 1.07 & 1.07 & 1.07\\
     $\kappa_1^N/\kappa_1^{N-1}$  & -  &      & 2.46 & 1.68\\
    \hline
    $\Gamma_0^N$ &  0.93 & 1.63 & 2.34 & 3.10 \\
    $\Gamma_1^N$ & - & 0.98 & 1.92 & 2.77 \\
  \end{tabular}
\end{center}
  \caption{We report the parameters $\kappa_n^N$, in unit of $b$, $\Gamma_n^N$,
  and selected ratios for the P\"oschl-Teller potential.}
  \label{tab:datiTP}
\end{table}

\begin{figure}[h]
    \includegraphics[width=\linewidth]{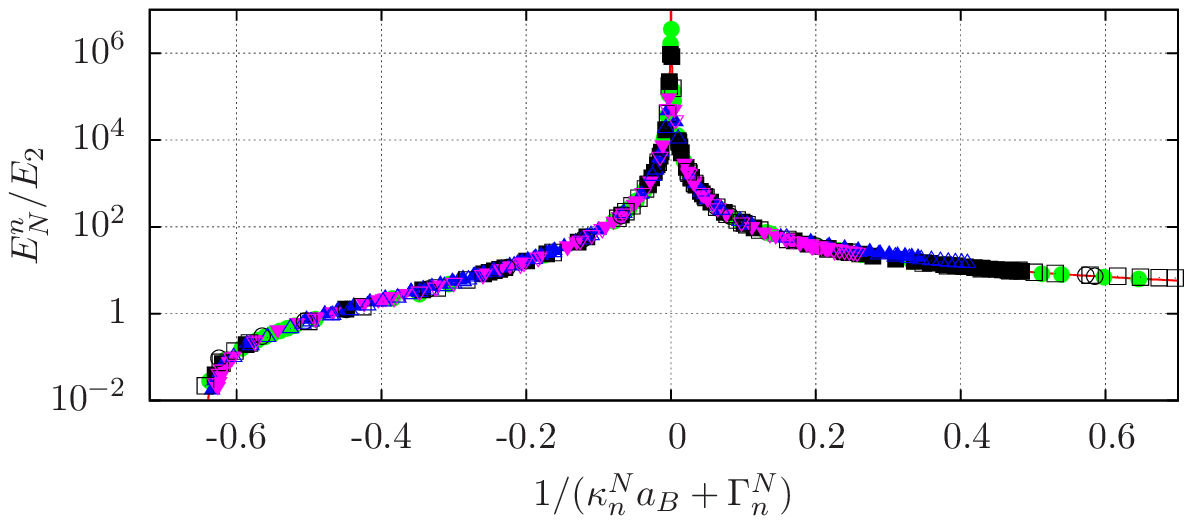}
    \caption{(color online). Ground- and excited-state N-body
      binding-energies $E^{n}_N$ in units of $E_2$ as a function of
      $1/(\kappa_{n}^N a_B +\Gamma_{n}^N)$.  $E_2$ is the two-body binding energy
      for $a>0$, and the two-body virtual state for $a<0$. The symbols are the
      same as in Fig.~2 and Fig.~3 of the paper. All the data collapse
      on the three-body universal curve (solid line) calculated 
      in the scaling limit.}
  \label{fig:universality}
  \end{figure}

In Fig.~\ref{fig:universality} we report all the calculated data, both with
Gaussian and PT potential.

\begin{figure}[h]
    \begin{center}
      \includegraphics[width=\linewidth]{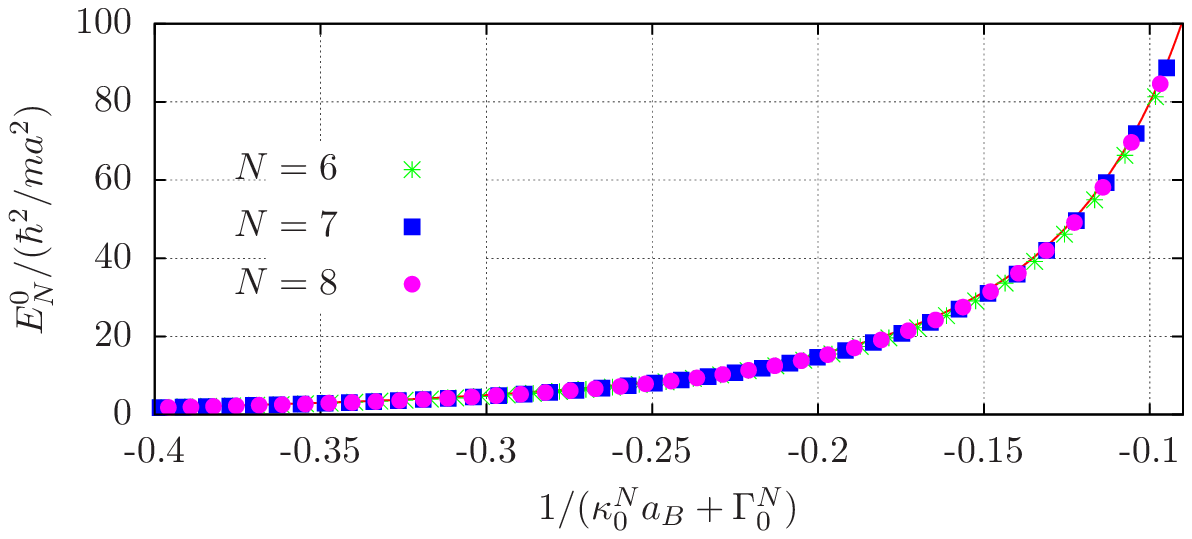}
    \end{center}
    \caption{The ground-state N-body binding-energy
      $E^0_N$  from Ref.~\cite{von_stecher:2010_j.phys.b:at.mol.opt.phys.}   in
    units of $\hbar^2/ma^2$ as a function of $1/(\kappa_0^N a + \Gamma_0^N)$.
  The solid curve represent the universal zero-range curve.}
  \label{fig:vonStecher}
  \end{figure}

\subsection{Analysis of published calculations of [J. von Stecher, J. Phys. B: At.
  Mol. Opt. Phys. 43, 101002 (2010)].}

We have applied Eq.(5) of our paper to analyse the calculation of 
Ref.~\cite{von_stecher:2010_j.phys.b:at.mol.opt.phys.}. In that paper the author 
performed few-body calculations using two- plus three-body Gaussian potentials for
$N=6$, and a two-body square-well  potential plus three-body hard-wall potential
for $N=7,8$. In that paper the author gives a four-parameter parametrization of
the ground-state energies for $N=6,7,8$, and we have used that parametrization
in order to reconstruct the calculated data.

\renewcommand{\arraystretch}{1.5}
\begin{table}[h]
  \begin{tabular}{ c | c c c }
    \hline
    \hline
     & $N=6$ & $N=7$  & $N=8$ \\
     \hline
     $\kappa_0^N /\kappa_0^3$ &  4.29 & 5.22 & 6.18\\
    $\Gamma_0^N$ & -0.393 & -0.414 & -0.382  \\
  \end{tabular}
  \caption{We report the parameters $\kappa_0^N$, in unit of $\kappa_0^3$, and $\Gamma_n^N$,
  used to analysis data of
  Ref.~\cite{von_stecher:2010_j.phys.b:at.mol.opt.phys.}.}
  \label{tab:vonStecher}
\end{table}

In Fig.~\ref{fig:vonStecher} we show that the extracted data, once analysed 
with Eq.(5) of our paper, collapse onto the universal zero-range curve.
As a side effect, we see that only two-parameters are needed to describe data, 
and that the
collapse does not depend on the Gaussian form of the potential. In
Table~\ref{tab:vonStecher} we report the parameters we used to collapse the
data.

\end{document}